\documentclass[useAMS,usenatbib]{mn2e}

\usepackage[T1]{fontenc}
\usepackage{aecompl}
\usepackage{graphicx}
\usepackage{overpic}
\usepackage{amsmath,amssymb}
\usepackage[draft]{hyperref}
\usepackage{color}

\title[Filamentation instability of pair plasmas]
{Particle acceleration and radiation friction effects in the filamentation instability of pair plasmas}

\author[M. D'Angelo et al]{M. D'Angelo$^{1}$\thanks{E-mail: marta.dangelo@gssi.infn.it (AVR)}, L. Fedeli$^{2,3}$, A. Sgattoni$^{3}$, F. Pegoraro$^{2}$ and A. Macchi$^{3,2}$ \\
$^{1}$Gran Sasso Science Institute-INFN, viale Francesco Crispi 7, L'Aquila, 67100, Italy \\
$^{2}$Dipartimento di Fisica Enrico Fermi, Universit\`a di Pisa, Largo Bruno Pontecorvo 3, Pisa I-56127, Italy \\
$^{3}$Istituto Nazionale di Ottica, Consiglio Nazionale delle Ricerche (CNR/INO), u.o.s. Adriano Gozzini, Pisa, Italy }      
                                                           
\begin{document}

\pagerange{\pageref{firstpage}--\pageref{lastpage}} \pubyear{2015}

\maketitle

\label{firstpage}


\begin{abstract}

The evolution of the filamentation instability produced by two counter-streaming, ultrarelativistic pair plasmas is studied with particle-in-cell simulations. Radiation friction effects are taken into account. Two dimensional simulations are performed for both cases of the initial momenta being perpendicular ($T$-mode) or parallel ($P$-mode) to the simulation plane. In the initial stage the instability is purely transverse for both modes and generates small-scale filaments which later merge into larger structures. Particle acceleration leads to a strong broadening of the energy spectrum with the formation of a peak at twice the initial energy for the $T$-mode. In the nonlinear stage significant differences between $T$- and $P$-modes in the evolution of the fields and in the spectra of accelerated particles are apparent. The presence of radiative losses does not change the dynamics of the instability but strongly affects the structure of the particle spectra in the ultra-relativistic regime (particle energy $>100~\mbox{MeV}$) and for high plasma densities ($>10^{21}~\mbox{cm}^{-3}$).
\end{abstract}



\section{Introduction}

From the 1970s on, the long-standing problem of high-energy cosmic ray origin has involved beam-plasma instabilities in order to explain some aspects of the acceleration mechanism (see~\citet{blandfordApJ78,bellMNRAS78I,bellMNRAS78II} or \citet{blasiAAPR13} for a more recent review). In particular the excitation of unstable modes by the accelerated particles propagating into the interstellar medium has been studied as a possible mechanism to generate strong magnetic turbulence predicted by the non-linear diffusive shock acceleration theory (see the reviews by \citet{malkovRPP01} and \citet{blandfordPR87}). 
The study of a model problem characterized by two countestreaming electron-positron plasma clouds at relativistic energies can be relevant to various astrophysical scenarios including the fireball model of Gamma Ray Bursts~\citep{piranRMP05}, pulsar wind outflows in Pulsar Wind Nebulae~\citep{blasi-amato11}, and relativistic jets from Active Galactic Nuclei~\citep{begelmanRMP84}. 
In this context, several authors have studied counterstraming pair plasmas in various configurations (see e.g. \cite{hoshinoApJ02,silvaApJ03,jaroschekApJ05,changAJ08,spitkovskyAJL08,amanoPoP09,nishikawaAPJ09,bretPoP13,liangApJL13a,liangApJL13b,lemoineMnras14}), including colliding and injected jets (as opposed to uniform configurations) which allow the generation of collisionless shocks. 
\newline
In this paper we examine the instability generated by two counter-streaming neutral beams of pair plasmas in the ultra-relativistic regime.  
In particular we address kinetic effects, such as particle acceleration, taking place during the nonlinear phase of the instability, and we take radiation friction (RF) effects into account. It is worth noticing that there is a current interest in kinetic simulations of pair plasmas with RF included in astrophysics, e.g. for the study of anomalous particle acceleration leading to flaring in the Crab Nebula \citep{jaroschekPRL09,ceruttiApJ13}. The problem of RF inclusion in the kinetic modeling of a relativistic plasma in high electromagnetic (EM) fields is also crucial in the context of ultraintense laser interaction with matter and plasma \citep[and references therein]{dipiazzaRMP12}. It is therefore useful to revisit classic plasma instabilities in highly relativistic regimes possibly dominated by radiation.  \newline
The system composed by two charge-neutral, counterstreaming pair plasmas is subject to a host of instabilities which depend on the orientation of the wavevector with respect to the direction of the beams (for a general review see~\citet{bretPOP10}). The unstable spectrum includes two limiting cases: the longitudinal two stream instability (TSI), corresponding to an electrostatic mode with flow-aligned wavevector, and the transverse filamentation instability (FI), corresponding to an EM mode with wavevector perpendicular to the beam direction. The TSI and FI are particular cases of the more general instability in which the wavevector is oblique to the beam direction and the unstable spectrum presents both the electrostatic and the EM components.
From analytical calculations based on first-order perturbation theory (\citet{califanoPRE97,kazimuraAJL98,bretPRE04,bretPOP10}), the growth rate $\Gamma$ in the linear phase for the two-stream instability $\Gamma \propto \gamma_0^{-3/2}$, while for the FI $\Gamma \propto \gamma_0^{-1/2}$, where $\gamma_0 = \left({1 + (p_0/m_e c)^2}\right)^{1/2}$ is the initial beam Lorentz factor (with $p_0$ the initial drift momentum). Moreover, these calculations show that, when the beams are symmetric, the instability is prevalently EM. Thus, in the ultra-relativistic regime the transverse FI is expected to dominate the growth of the instability, at least before nonlinear effects become important.\newline

We performed EM, fully relativistic {particle-in-cell} (PIC) simulations both in one spatial dimension (1D) and in two spatial dimensions (2D) with plane Cartesian geometry. In 2D, the simulations can be performed with either the counterstreaming beams direction perpendicular to the simulation plane ($T$-mode) or parallel to it ($P$-mode). For the $T$-mode case, only the growth of transverse modes is allowed, while the $P$-mode allows longitudinal modes as well. Thus, in general the dynamics of countestreaming instabilities in 2D can be substantially different between $T$- and $P$-modes (see e.g. \citet{amanoPoP09} for the case of Kelvin-Helmoltz instability in electron-ion plasmas and \citet{liangApJL13a,liangApJL13b} for shear instability in pair plasmas) so that in principle a three-dimensional (3D) analysis would be needed. However, a reliable 3D simulation is often not possible because of the huge computational cost, which leads to severe limitations in the numerical box size, spatial and temporal resolution, and number of particles per cell even on a parallel supercomputer. This is particularly true for our study where we aim at understanding kinetic and particle acceleration effects, which need sufficient phase space statistics, i.e. large number of computational particles. A similar request holds in order to address the effects of the RF force, since the latter is much stronger on the highest energy particles in the low-density tail of the particle distribution.  In addition, the strong coalescence of small scale structures in the nonlinear stage eventually leads to the formation of structures with size close to the numerical box. For these reasons, a ``small'' 3D simulation would excessively suffer from numerical effects at present. Therefore in this paper we consider only 2D simulations, assuming that a comparison of $T$- and $P$-mode simulations can give insight into the 3D dynamics. We restrict to a configuration of homogeneous, counterstreaming plasmas which prevent the formation of shocks, that are not of direct interest for this paper. However it should be noticed that the nonlinear dynamics and saturation of the instability may be different for colliding or injected jets configuration.
\newline

Although as it will be shown below the transverse mode is dominant in the early, linear stage leading to the generation of filaments (which are actually current layers in the $P$-mode), significant differences between the $T$- and $P$-mode appear in the nonlinear phase. In both cases, the transition from the linear to the nonlinear phase is characterized by the coalescence of the current filaments, with progressive decay of the magnetic field after reaching a peak value at the endo of linear phase. Differences between the $T$- and $P$-mode appear in the nonlinear phase, with the amplitude of the magnetic field at peak and at late times being stronger for the $T$-mode. In addition, particle spectra are significantly different, with the formation of a spectral peak for the $T$-mode only, while the high energy cut-off is higher for the $P$-mode. Species separation is also different between $T$- and $P$-modes. The high energy tail of the particle spectrum is strongly affected by RF effects, which however do not cause substantial modifications in the dynamics of instability and in the temporal evolution of fields.


\section{Simulation model}

\subsection{Numerical set-up}
\label{setup}

The initial configuration of our simulations consists of two neutral beams of electron-positron pairs which propagate in opposite directions (corresponding to $p_z$ in the momentum space) and fill the entire simulation box. 
The system is symmetric, with the populations of the two beams having the same initial density, i.e. $n_{e1}^{(0)} = n_{e2}^{(0)} = n_{p1}^{(0)} = n_{p2}^{(0)} = n_T/4$, where $n_T$ is the total density, and the same momentum absolute value, i.e. $p_{e1}^{(0)} = p_{e2}^{(0)} = p_{p1}^{(0)} = p_{p2}^{(0)} = p_0$, and consistently the initial values of charge and current densities and of the electric and magnetic fields are zero. A very small temperature is introduced to seed the instability. In both 1D and 2D cases we used periodic boundary conditions.\newline
We performed simulations with different Lorentz factors $\gamma_0$ from 1 to $10^3$. Here we describe the case with $p_0/m_e c = 200$ as it is representative of the most relevant effects observed.
For different values of $p_0$, there are no qualitative changes in the dynamics of the instability, the most important difference being the growth rate of the modes which scales as $\gamma_0^{-1/2}$ (see e.g. \citet{califanoPRE97,kazimuraAJL98,bretPRE04}). \newline
In the $1$D case, the simulation box is aligned along the $\mathbf{x}$-direction and it is divided into $15000$ grid cells of equal length $\Delta x = 0.01 \, \lambda_p$ with $\lambda_p=c/\omega_p$ the skin depth and $\omega_p = \left({4 \pi e^2 n_T/m_e}\right)^{1/2}$. Each of the four plasma species is represented by $N_p = 3 \times 10^{6}$ computational particles ($200$ particles per cell). The total simulation time is $t_{sim} = 1000 \, T_p$, where $T_p = 2 \pi/ \omega_p$, and the temporal resolution is $\Delta t = 0.01 \, T_p$. \newline
For the 2D $T$-mode case, the open-source code PICCANTE \citep{PICCANTE,PICCANTE2}, optimized for parallel simulations, has been used. In this case the box had $2000 \times 2000$ cells and lengths $L_{g,x} \times L_{g,y} = 100 \, \lambda_p \times 100 \, \lambda_p$ so $\Delta x = \Delta y = 0.05 \, \lambda_p$. For each species $N_p=2 \times 10^8$ computational particles and $t_{sim} = 200 \, T_p$ with $\Delta t = 0.0325 \, T_p$.  \newline
For the $P$-mode, simulations performed using the standard Finite Difference Time Domain (FDTD) Maxwell solver algorithm of PICCANTE were strongly affected by numerical \v{C}herenkov radiation (NCR; for details see \citet{greenwoodJCP04}) due to high-frequency waves which propagate slower than high-energy particles. Thus, for the $P$-mode simulations we set up another PIC code (PICcolino) implementing a spectral Maxwell solver based on the Fast Fourier Transform, which is free from NCR. PICcolino was benchmarked with PICCANTE in a series of cases where NCR was negligible, e.g. in $T$-mode simulations, showing full agreement in the results. The only noticeable difference was some time delay in the early rise of the instability (but with the same growth rate) due to a slightly different level of initial noise. For $P$-mode simulations with PICcolino, the box had $1000 \times 1000$ cells wtih $L_{g,x} \times L_{g,z} = 50 \, \lambda_p \times 50 \, \lambda_p$, $\Delta x = \Delta z = 0.05 \, \lambda_p$, $N_p=5 \times 10^7$, $t_{sim} = 200 \, T_p$ and $\Delta t = 0.025 \, T_p$. 

\subsection{Radiation friction modeling}
\label{sec:RFM}

The inclusion of RF in the code is based on the Landau-Lifshitz approach \citep{landauRR}, with the approximations and the numerical implementation introduced by \citet{tamburiniNJP10}; see also \citet{vranicXXX15} for a further discussion and comparison to other approaches.  
The radiation friction force which acts on the particles in addition to the Lorentz force is
\begin{eqnarray}
\mathbf{f}_{RF}
&=&-\frac{2}{3}r_e^2\left(\gamma^2\left(\left(\mathbf{E}+\frac{\mathbf{v}}{c}\times\mathbf{B}\right)^2-\left(\frac{\mathbf{v}}{c}\cdot\mathbf{E}\right)^2\right)\frac{\mathbf{v}}{c}\right.\nonumber \\ 
& &\left. - \left(\mathbf{E}+\frac{\mathbf{v}}{c}\times\mathbf{B}\right)\times\mathbf{B}-\left(\frac{\mathbf{v}}{c}\cdot\mathbf{E}\right)\mathbf{E}\right)
\, ,
\label{eq:RF}
\end{eqnarray}
where $r_e \equiv e^2/m_e c^2 \approx 2.8 \times 10^{-9} \mathrm{\mu m}$ is the classical electron radius and $\lambda = 2\pi\lambda_p$. A third term in the Landau-Lifshitz expression has been neglected since it is negligible in all situations where use of $\mathbf{f}_{RF}$ is appropriate. 
In the ultra relativistic regime the most important contribution in Eq.~\eqref{eq:RF} comes from the first term because it is proportional to particle Lorentz factor $\gamma^2 \gg 1$. The numerical implementation in the PIC code is discussed by \citet{tamburiniNJP10}. 
The Compton drag force is neglected.
\newline
In the case without RF inclusion, the equations of the PIC code are in an universal dimensionless form with the density normalized to $n_T$, time to $1/\omega_p$, space to $c/\omega_p$, and fields to $m_ec\omega_p/e$. Thus, all the results of a simulation can be scaled with respect to a definite value for the density. The inclusion of RF breaks such scaling invariance, so it is necessary to set a dimensional value for the plasma density. We have performed simulations with RF included for density values up to $n_T = 10^{21} \mathrm{cm^{-3}}$. \newline
The friction effect of the RF force physically arises from the incoherent emission of high-frequency radiation by ultra-relativistic electrons and positrons, see \citet{dipiazzaRMP12}. From a numerical point of view, it is unfeasible to perform simulations with a spatial resolution high enough to resolve such a small wavelength radiation. Thus, it is assumed that such radiation escapes from the system without re-interacting with other electrons or positrons, and the RF acts as a loss term. For density values of the order $\leq 10^{21} \mathrm{cm^{-3}}$, it can be safely assumed that the plasma is optically thin to the high-frequency radiation (having a typical energy $\ge \mathrm{MeV}$) which mostly contributes to radiation losses. In addition, the mean free path  for Compton scattering of photons is $\ge 1 \, \mathrm{m}$ (as estimated from the Klein-Nishina formula), typically much larger than the scale length on which the instability sets up (of the order of $l_p=c/\omega_p \sim \mathrm{\mu m}$).  \newline


\subsection{Symmetry relations}
\label{sec:fluid}

In a cold four-fluid description, the transverse unstable mode exhibits symmetry properties whose violation is a signature for kinetic and nonlinear effects. Let us indicate the density of particles having $p_0 > 0$ with ${n}_{\rightarrow}^{\pm}$ for positrons and electrons, respectively, and similarly we define $n_{\leftarrow}^{\pm}$ for particles having $p_0<0$. We use the same notation for all other fluid variables. For fields and gradients  we use $\parallel$ and $\perp$ to indicate quantities parallel and perpendicular to the beams, respectively. For the EM transverse unstable mode with wavevector $\mathbf{k}$ the electric field is parallel to the beams ($\mathbf{E} = (0 \, ,0 \, ,E_z) = \mathbf{E}_{\parallel}$)  while $\mathbf{B} = (B_x \, , B_y,0) = \mathbf{B}_{\perp} \perp \mathbf{k}$ (we neglect the effect of transverse components of $\mathbf{E}$ which do not play a role in the linear stage of the instability but might be generated due to nonlinear charge separation effects). The EM field can be thus described via a vector potential $\mathbf{A} = (0 \, ,0 \, ,A_z) = \mathbf{A}_{\parallel}$ such that $\mathbf{E}_{\parallel} = - \partial_t \mathbf{A}_{\parallel}/c$ and $\mathbf{B}_{\perp} = \nabla_{\perp} \times \mathbf{A}_{\parallel}$ (we note that $\nabla_{\perp} = (\partial_x \, , \partial_y, 0) = \nabla$).  The fluid equations can be thus written as
\begin{align}
&\partial_t n_{\leftrightarrow}^{\pm}+\nabla_{\perp} \cdot \left(n_{\leftrightarrow}^{\pm}\mathbf{u}_{\perp, \leftrightarrow}^{\pm}\right)=0 \, ,\label{eq:fluid1}\\
&D_{t,\leftrightarrow}^{\pm}\left(\gamma_{\leftrightarrow}^{\pm} \mathbf{u}_{\perp,\leftrightarrow}^{\pm}\right)=\mp (e/m_ec)u_{z,\leftrightarrow}^{\pm}\nabla_{\perp} \times \mathbf{A}_{\parallel} \, ,\label{eq:fluid2}\\
&D_{t,\leftrightarrow}^{\pm}\left(\gamma_{\leftrightarrow}^{\pm}\mathbf{u}_{\parallel,\leftrightarrow}^{\pm}\right)=\mp(e/m_ec)D_{t,\leftrightarrow}^{\pm} \mathbf{A}_{\parallel} \, ,\label{eq:fluid3}
\end{align}
where $D_{t,\leftrightarrow}^{\pm}\equiv\left(\partial_t+\mathbf{u}_{\leftrightarrow}^{\pm}\nabla\right) = \left(\partial_t+\mathbf{u}_{\perp,\leftrightarrow}^{\pm}\nabla_{\perp}\right)$. The vector potential satisfies the wave equation
\begin{equation}
\left(\nabla^2 - c^{-2}\partial_t^2\right)\mathbf{A}_{\parallel} = - \frac{4\pi}{c} \mathbf{J}_{\parallel} \; ,
\end{equation}
where 
\begin{align}
 \mathbf{J}_{\parallel} 
= &+e\left({n}_{\leftarrow}^+\mathbf{u}_{\parallel,\leftarrow}^++{n}_{\rightarrow}^+\mathbf{u}_{\parallel,\rightarrow}^+\right)
\nonumber \\
&-e\left({n}_{\leftarrow}^-\mathbf{u}_{\parallel,\leftarrow}^-+{n}_{\rightarrow}^-\mathbf{u}_{\parallel,\rightarrow}^-\right) \; .
\end{align}
The system is symmetric under the transformation that reverses at the same time the charge and the direction of propagation of the populations, which simplifies the description of the dynamics. It is also possible to reduce the initial set of equations to a system involving only two populations and three pairs of dynamical variables, as done in \citet{kazimuraAJL98}. The two populations are the sources of the positive ($\mathbf{J}_{\parallel}^+$) and negative ($\mathbf{J}_{\parallel}^-$) density current and are identified with $+$ and $-$ symbols. The two-fluid variables are defined as follows: $ n^{+} = n_{\rightarrow}^+ + n_{\leftarrow}^-$, $n^{-} = n_{\leftarrow}^+ + n_{\rightarrow}^-$, $\mathbf{u}_{\perp}^{+} = \mathbf{u}_{\perp,\rightarrow}^+ = \mathbf{u}_{\perp,\leftarrow}^-$, $\mathbf{u}_{\perp}^{-} = \mathbf{u}_{\perp,\leftarrow}^+ = \mathbf{u}_{\perp,\rightarrow}^-$, $\mathbf{J}_{\parallel}^{+} = e (n_{\rightarrow}^+\mathbf{u}_{\parallel,\rightarrow}^+ - n_{\leftarrow}^- \mathbf{u}_{\parallel,\leftarrow}^-)$ and $\mathbf{J}_{\parallel}^{-} = e (n_{\leftarrow}^+\mathbf{u}_{\parallel,\leftarrow}^+- n_{\rightarrow}^-\mathbf{u}_{\parallel,\rightarrow}^-)$. 
The two-fluid system of equations which is obtained from Eqs.(\ref{eq:fluid1}-\ref{eq:fluid3}) is
\begin{align}
 \partial_t n^{\pm} + \nabla_{\perp} \cdot (n^{\pm}\mathbf{u}_{\perp}^{\pm}) = 0 \; , \\
 \partial_t(n^{\pm}\mathbf{u}_{\perp}^{\pm}) + \nabla_{\perp} \cdot[n^{\pm}\mathbf{u}_{\perp}^{\pm} \otimes\mathbf{u}_{\perp}^{\pm}] = \frac{1}{m_ec}\mathbf{J}_{\parallel}^{\pm} \times(\nabla_{\perp} \times \mathbf{A}_{\parallel}) \; , 
\\
 [\partial_t + \nabla_{\perp} \cdot \mathbf{u}_{\perp}^{\pm}](\gamma^{\pm}\mathbf{J}_{\parallel}^{\pm}) 
 = -\frac{e^2}{m_ec}n^{\pm}[\partial_t+ \mathbf{u}_{\perp}^{\pm}\cdot\nabla_{\perp}] \mathbf{A}_{\parallel}  \; . 
\end{align}

\section{Simulation results}

\subsection{One-dimensional simulations}
\label{sec:1D}

We first study the FI in 1D mostly as a test bed and guidance for multi-dimensional simulations. The 1D model has the advantages of being directly comparable to analytical results, and in particular to check symmetry properties and conservation laws. In addition, the 1D geometry allows high resolution runs and detailed analysis of simulation data. We performed several tests changing the number of particles per cell and the spatial resolution in order to check the sensitivity of the results. 
\newline     
\begin{figure}
\begin{overpic}[width=0.5\textwidth]{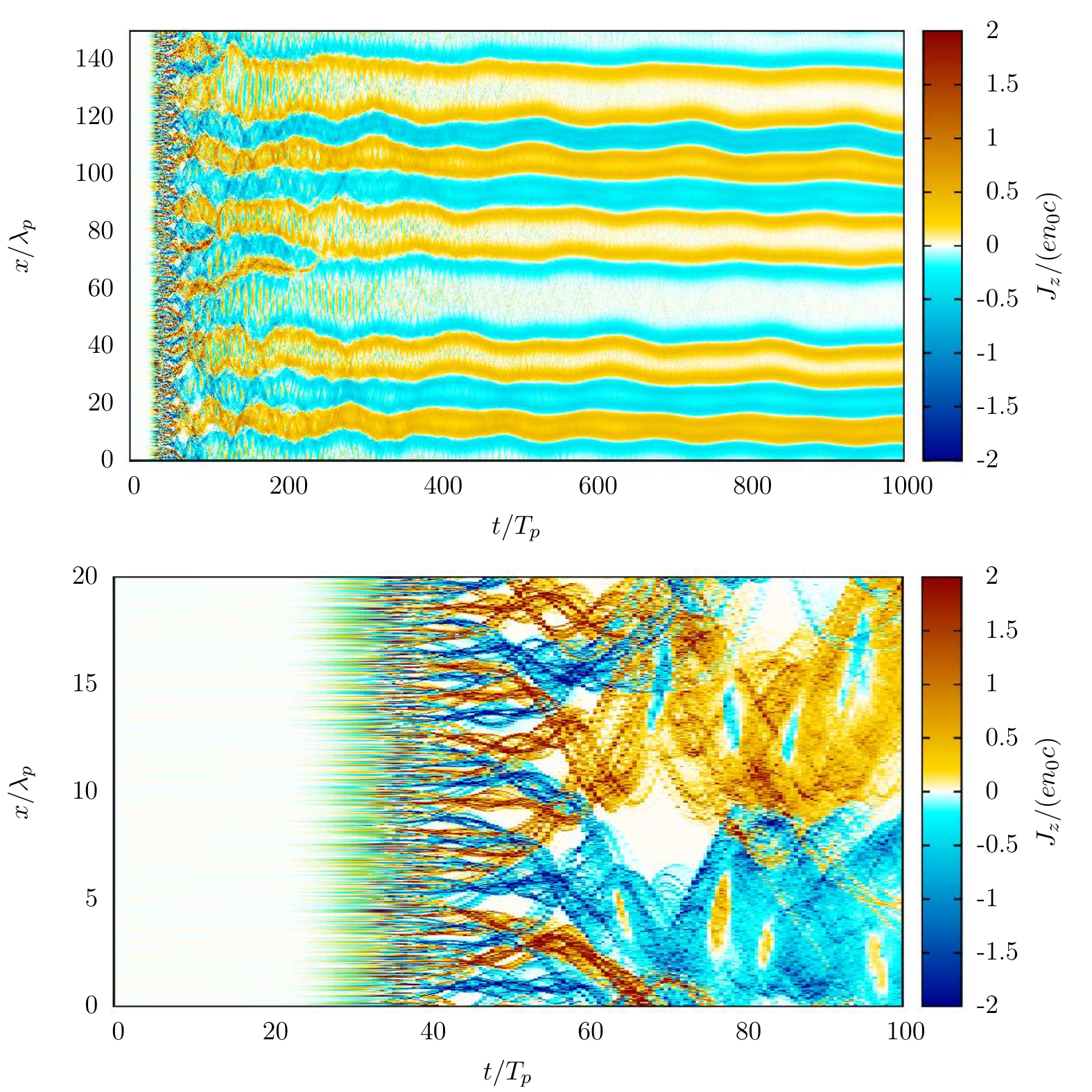}
\put(5,98){$(a)$}
\put(5,51){$(b)$}
\end{overpic}
   \caption{1D simulation: the current density $J_z$ as a function of position ($x$) and time ($t$). The upper plot, panel (a), shows the complete evolution over the whole simulation box, while the lower plot, panel (b), focuses on the early stage with small-scale filaments. The parameters $\lambda_p=c / \omega_p$ and time $T_p=2\pi/\omega_p$.}
   \label{fig:Jzpvt}
\end{figure}
The structure of the current density $J_z$ as a function of $(x,t)$ is shown in Fig.~\ref{fig:Jzpvt}~(a).  
The development of the instability can be divided into three phases: a linear phase for $t < 50 \, T_p$, a transition phase for $ 50 \, T_p < t < 200 \, T_p $ and a nonlinear, quasi-stationary phase for $t > 200 \, T_p$. 
In the linear phase, modes with a defined wavevector grow exponentially, as we verified by calculating the spatial Fourier transform $\mathcal{F}[B_y] = \hat{B}_y(k_x,t)$. The numerically obtained growth rate for every mode $k_x$ agrees well with analytic calculations \citep{kazimuraAJL98}.
\newline
In Fig.~\ref{fig:Jzpvt}~(b) a zoom on the structure of $J_z$ during the linear and the transition phase is shown.
During the exponential growth of the perturbations, $J_z$ has a filamentary structure with a very small scale length ($\ll \lambda_p$). At $t \sim 50 \, T_p$ separate filaments of opposite current, having a typical scale close to the electron skin depth, become distinguishable and start to merge. This coalescence characterizes the transition of the instability from the linear to the nonlinear quasi-stationary regime.
For $t > 200 \, T_p$ the merging phase of the filaments finishes and the size of each filament is constant, so that the configuration can be described as stationary except for some ``vibration'' which is observable in Fig.~\ref{fig:Jzpvt}~(a). \newline
\begin{figure}
    \includegraphics[width=0.5\textwidth]{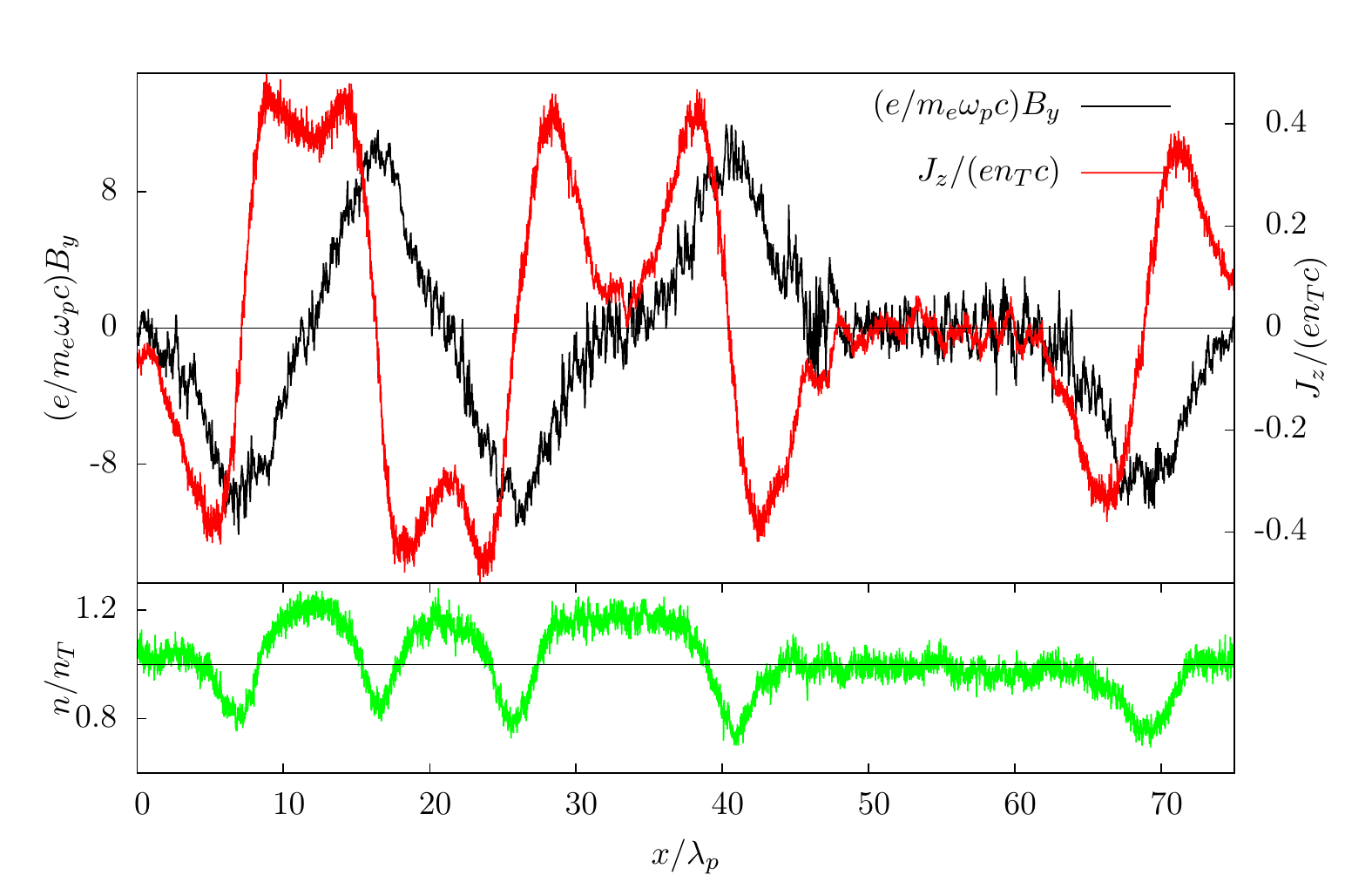}
    \caption{1D simulation: spatial profile of $J_z$ (red line), $B_y$ (black line) and $n$ (total number density, green line) at $t = 700 \, T_p$. Two adjacent maxima or minima of $J_z$ identify a positive or negative current filament, respectively.}	
   \label{fig:ByJz70000}
\end{figure}
To understand in more detail the nonlinear regime we consider the spatial profile of $J_z$, $B_y$ and $n$ (total number density), at $t = 700 \, T_p$, reported in Fig.~\ref{fig:ByJz70000}. 
A filament with positive or negative current is identified by two consecutive maxima or minima, respectively. Within each positive filament, the current density assumes its maximum value near the edges. Moving towards the inner region of the filament $J_z$ decreases assuming a local minimum at the center, whereas the total number density becomes flat-top. The same happens for negative current filaments. This feature corresponds to an anti-correlation between particle density and velocity, which will be further discussed below by looking at phase space distributions. 
An oscillatory pattern characterizes also the profile of the magnetic field $B_y$, which has null points at the center of each filament, as it is shown in Fig.~\ref{fig:ByJz70000}.  \newline
In the late, quasi-stationary phase the spatial structures of $J_z, \, B_y$ and $n$ indicate an accumulation of particles within the current filaments due to magnetic trapping. 
In this phase the characteristic scale length of the field becomes comparable to the Larmor radius and the density of the magnetic energy is of the order of the initial energy density:
\begin{equation}
 \frac{B_{sat}^2}{8 \pi} = n_T \gamma_0 m_e c^2 \, .
 \label{eq:equip}
\end{equation}
From Eq.~\ref{eq:equip} we estimate the Larmor radius as $r_{L,sat} =  \sqrt{\gamma_0/2} \, \lambda_p$, which gives the scale length of a filament $ r_f \approx r_{L, sat} \sim 10 \, \lambda_p$, in agreement with the numerical results (see Fig.~\ref{fig:ByJz70000}).   \newline
\begin{figure}
      \begin{overpic}[width=0.5\textwidth]{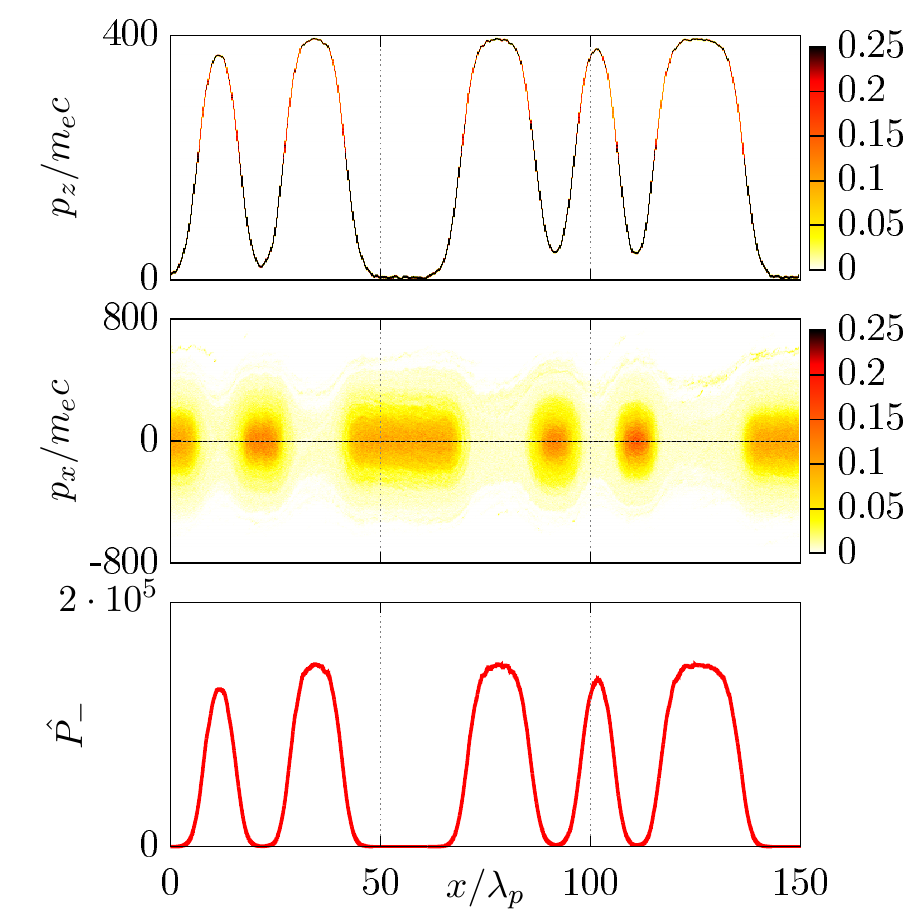}
      \put(0,95.5){$(a)$}
     \put(0,65){$(b)$}
     \put(0,30){$(c)$}
     \end{overpic}
       \caption{1D simulation: (a) phase space contours $(x,p_z)$; (b) phase space contours $(x,p_x)$; (c) effective potential $\hat{P}_{-}(x) = (p_0+a_z)^2$, defined in the eq.~\eqref{eq:effectivepot}. Plots (a) and (b) show the behavior of the electrons with $p_0/m_e \, c = 200$. All these quantities are plotted at $t = 800 \, T_p$. The color bar indicates number density.}  
       \label{fig:effectivepot}
\end{figure}
Figures~\ref{fig:effectivepot}~(a) and~(b) represent, respectively, the projection of the phase space on the $(x,p_z)$ and the $(x,p_x)$ planes for the electrons with initial positive momentum at $t = 800 \, T_p$, i.e. described by fluid variables $f^{-}_{\rightarrow}$. 
The momentum $p_z$ is a single valued function of the position $x$, so that we may also speak of $p_z$ as a well-defined quantity in fluid equations. In the regions of peak density, i.e. in the inner part of each filaments, $p_z \approx 0$, which is consistent with the local minimum of the current density that peaks at the edge of the filament. Outside the filaments there is a small number of electrons which have $p_z \approx 2 \, p_0 = 400$. 
The $(x, p_x)$ phase space projection shows a spread along the longitudinal momentum $p_x$ with an approximately Gaussian distribution. \newline 
Consistently with the symmetry properties of the system (see Sec.\ref{sec:fluid}), the positrons with initial negative momentum, described by $f^{+}_{\leftarrow}$, have the same spatial distribution as the $f^{-}_{\rightarrow}$ electrons. Thus we may also consider Fig.~\ref{fig:effectivepot} as being representative of the $f^{+}$ population in the two-fluid description. The particles of the $f^{-}$ population show a pattern analogous to Fig.~\ref{fig:effectivepot} with their spatial distribution in space being complementary to that of the $f^{+}$ population, i.e. corresponding to oppositely directed current filaments.\newline
In the nonlinear, quasi-stationary regime the spatial distribution of particles may be described in terms of an effective potential as follows. 
The conserved canonical momentum is
\begin{equation}
 \Pi_z = p_z \pm a_z \; ,
 \label{eq:canmom}
\end{equation}
where ${a}_z = (e/m_e c^2){A_z}$ is the dimensionless vector potential and $\pm$ refers to the sign of the particle charge. 
At $t = 0$ we have ${a}_z = 0$, so $\Pi_{z} = \pm p_0$ for the two beams, respectively. The normalized energies (in units of $m_ec^2$) of particles belonging to the populations are given by 
\begin{equation}
{\cal E}_{\pm}^2 = 1 + p_{x}^2 + [\mp p_0 + a_z(x)]^2 
\equiv 1 + p_{x}^2 + \hat{P}_{\pm}(x)
\, . 
\label{eq:effectivepot}
\end{equation}
The asymptotic state of the system may thus be described as a state in which the particles cluster into the minima of the effective potential $\hat{P}_{\pm}(x)$. Fig.~\ref{fig:effectivepot}~(c) shows $\hat{P}_-(x)$ at $t = 800 \, T_p$.

\begin{figure}
\includegraphics[width=0.5\textwidth]{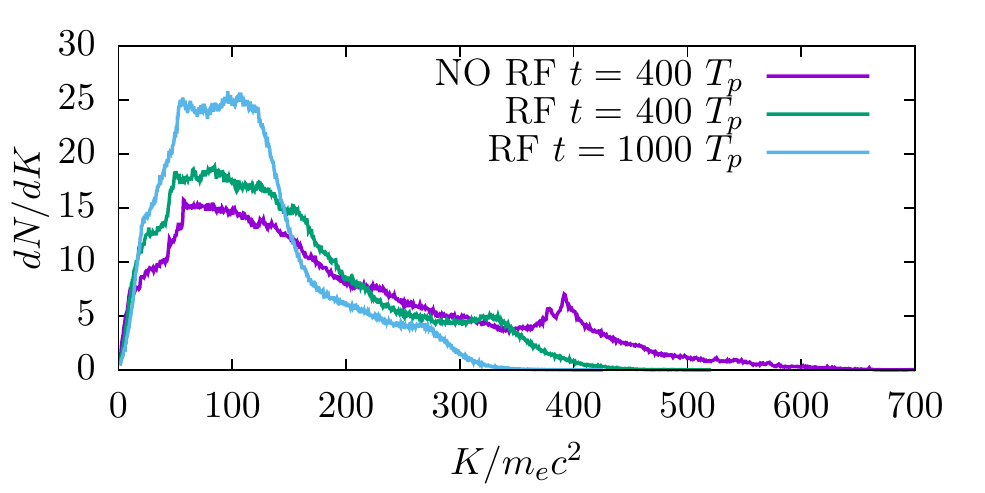}
     \caption{1D simulation: kinetic energy spectrum of the $f_{\rightarrow}^-$ population at different times, for simulations with and without RF. In the case without RF (shown at $t=400\,T_p$) a spectral peak appears at twice the initial beam energy $\simeq 2 \gamma_0 m_e c^2$, and the spectrum is almost unchanged at late times. For the case with RF, the spectral peak is smoothed out at $t=400\,T_p$ and at later times ($t=1000\,T_p$) the high energy tail is ``washed'' out because of radiative losses.
}
    \label{fig:enspec}
\end{figure}

Fig.~\ref{fig:enspec} shows the kinetic energy spectrum of the $f_{\rightarrow}^-$ population for different times, and for both cases in which RF is either included or not.
As expected from symmetry relations, the spectrum is essentially identical for the other three populations.
Without RF, the energy spectrum shows a sharp a peak at twice the initial kinetic energy, see Fig.~\ref{fig:enspec}. Correspondingly, we observe a sharp, peaked cut-off at $2p_0$ in the spectrum of $p_z$ (not shown).
The peak is strongly smoothed in the case with RF, which leads to cooling of the plasma by removing particles in the high energy tail, for which RF is much stronger due to the $\sim \gamma^2$ scaling. During the evolution of the system, RF effects on the particle spectra become more important in the nonlinear phase because of the generation of both strong magnetic fields (which lead to synchrotron emission) and the acceleration of particles to high energy. However, the early development of the instability and the structure and amplitude of the fields at saturation are weakly affected by RF.

The acceleration of particles which double the initial value of $p_z$ may be explained as follows. First we notice that for each of the two populations ($f^+$ and $f^-$) in the two-fluid description of the system, the high-energy particles having $p_z = \pm 2 \, p_0$ are localized outside the filaments where most of the particles belonging to the \emph{other} population are localized, as shown in Fig.~\ref{fig:effectivepot}~(a) and Fig.~\ref{fig:effectivepot}~(b).
In a given position $x$ where the field $E_z = E_z(x,t)$ acts on a species in such a way to reduce its initial momentum $p_0$, it necessarily acts on the counter-streaming species increasing its initial momentum. 
If a particle belonging to the $f^+$ population falls in a decelerating region for the $f^-$ population (i.e. a local minimum of the effective potential $\hat{P}_{-}$),  it gains the same momentum $p_z$ that is lost by the particles of the counter-streaming fluid. 
\newline

The acceleration mechanism may also be described using the effective potential, Eq.(\ref{eq:effectivepot}). For a particle belonging to one of the two fluid populations we have
\begin{equation}
  1 + p_{x}^2 + [\pm p_0 + a(x)]^2 = C_{\mp} \, , 
  \label{eq:conkin1}
\end{equation}
for the $f^{-}$ and $f^{+}$ fluids, respectively; $C_{\mp}$ are constants. Fig.~\ref{fig:effectivepot}~(a-b) shows than for a particle of the $f^{-}$ population $p_x^{-}$ has a maximum in positions $x_0$ where $|p_{z}^{-}(x_0)| = 0$, while in the same position $|p_{z}^{+}(x_0)|$ is maximum and $p_{x}^{+}(x_0) = 0$ for the $f^{+}$ population.
Thus $p_x^{+}$ assumes its maximum value at $x_0$ where $p_{z}^{-}(x_0) = 0$ i.e. $a(x_0) = -p_0$. Eq.~\eqref{eq:conkin1} then yields $C_{+} = 1 + 4 p_0^2$.
Due to the symmetry of the system, the vector potential $a(x)$ assumes the same values for its maxima and minima as a function of $x$, so there will be another point $x^{'}_0$ where $a(x^{'}_0) = p_0$, which using Eq.~\eqref{eq:canmom} yields the maximum value of the momentum $|p_{z}^{+}(x_0')| = 2 p_0$. For symmetry reasons we also obtain $C_{-} = 1 + 4p_0^2$ and a maximum of $2p_0$ for $|p_{z}^{-}|$. Thus the maximum energy of the particles is
\begin{equation}
{\cal E}_{\mbox{\tiny max}} = ({1 + 4 p_0^2})^{1/2} \; .
\end{equation}
The particles with maximum $p_z$ are in positions where $|a_z|$ has a maximum or minimum. Thus, $dp_z/dx = \pm da_z/dx = 0$ where $|p_z|=2p_0$, i.e. the particles all gain the same momentum to first order in their distance from the maximum of $|a_z|$, which explains the peak at the cut-off in the energy spectrum (Fig.~\ref{fig:enspec}) as the formation of a spectral caustic.   

%


\begin{figure*}
\includegraphics[width=1.0\textwidth]{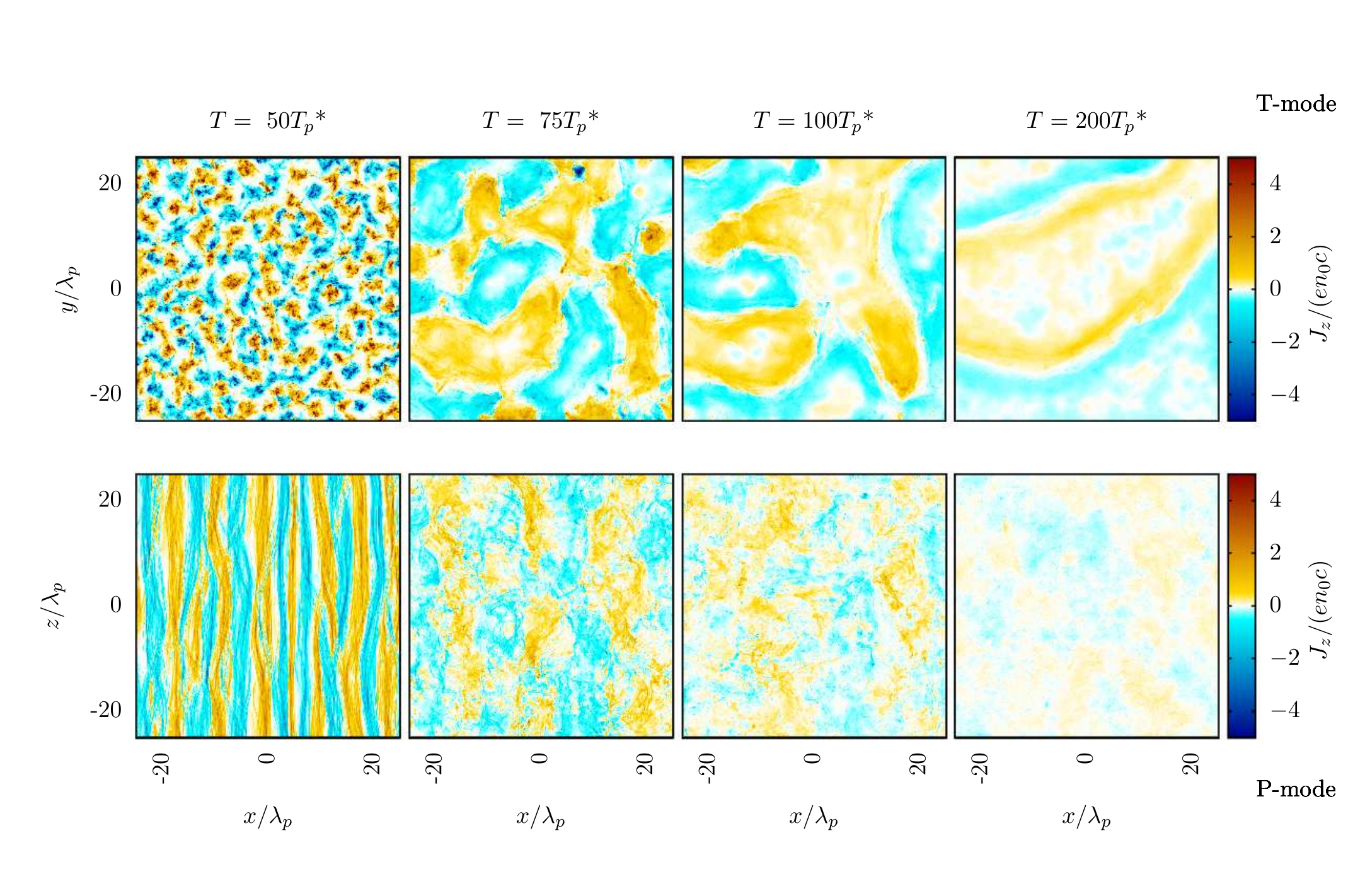}
\caption{Spatial distribution of $J_z$ in 2D simulations for the $P$-mode (lower row) and $T$-mode (upper row) cases, at four different times (simulation times have been shifted in order than the instant of peak magnetic energy coincides for the two cases, see text for details). Only a quarter of the whole box is plotted for the T-mode simulation.}
\label{fig:J2DPT} 
\end{figure*}

\subsection{Two-dimensional simulations}
\label{sec:2D}

\begin{figure*}
\includegraphics[width=1.0\textwidth]{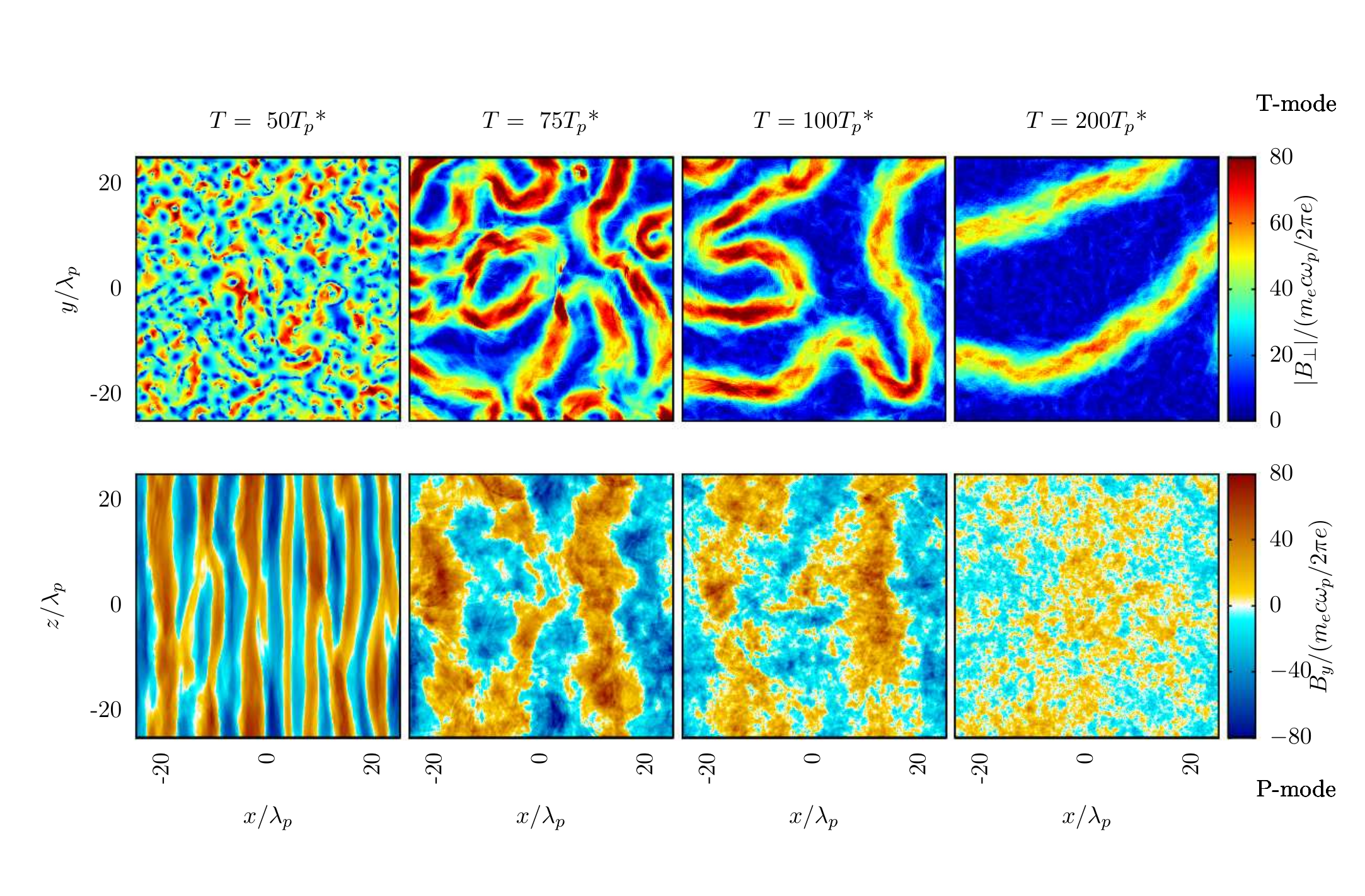}
\caption{Spatial distribution of the  magnetic field in 2D simulations at the same times of the distributions of $J_z$ in Fig.~\ref{fig:J2DPT}. For the $P$-mode case (loweer row) $B_y$ is shown. For the $T$-mode case (upper row) $|B_{\perp}|=(B_x^2+B_y^2)^{1/2}$ is shown. Only a quarter of the whole box is plotted for the T-mode simulation.}
\label{fig:B2DPT}
\end{figure*}

\begin{figure}
\includegraphics[width=0.5\textwidth]{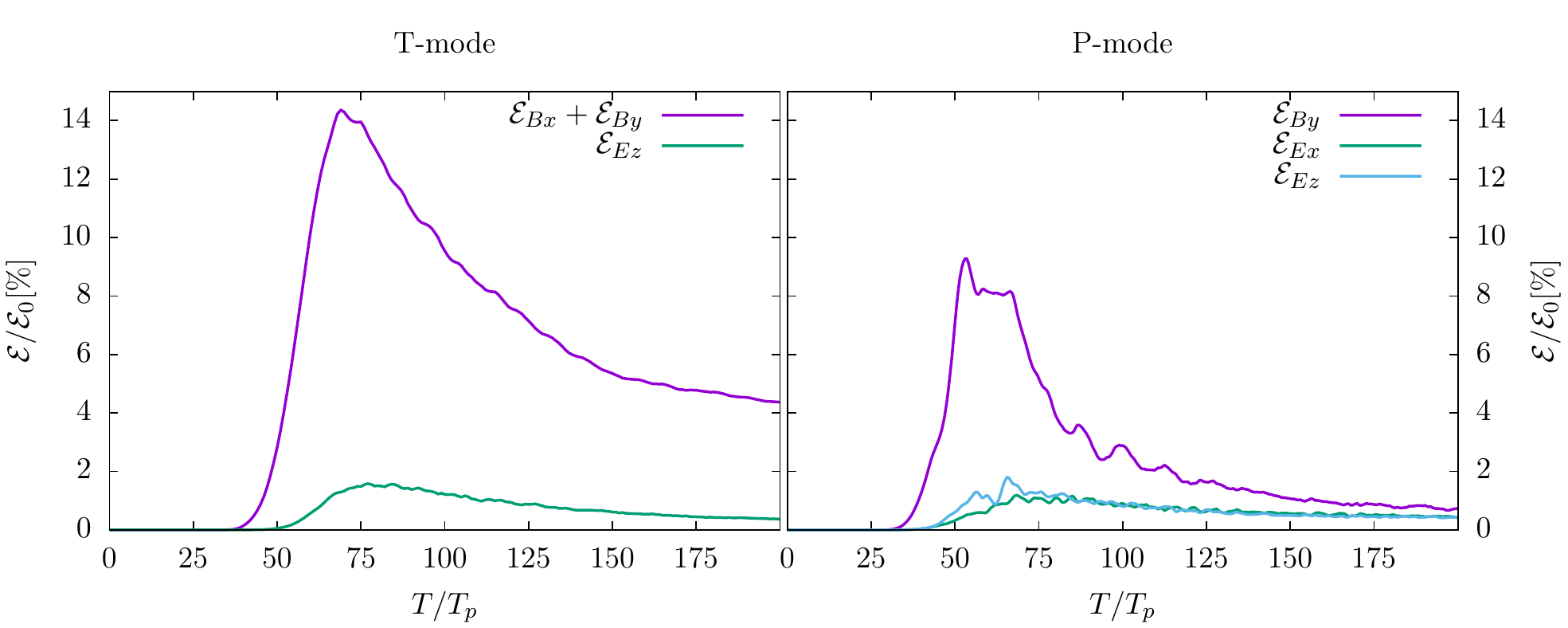}
\caption{Time history of field energy in 2D simulations for both $T$- and $P$-modes. }
\label{fig:energy2D}
\end{figure}

\begin{figure}
 \begin{overpic}[width=0.5\textwidth]{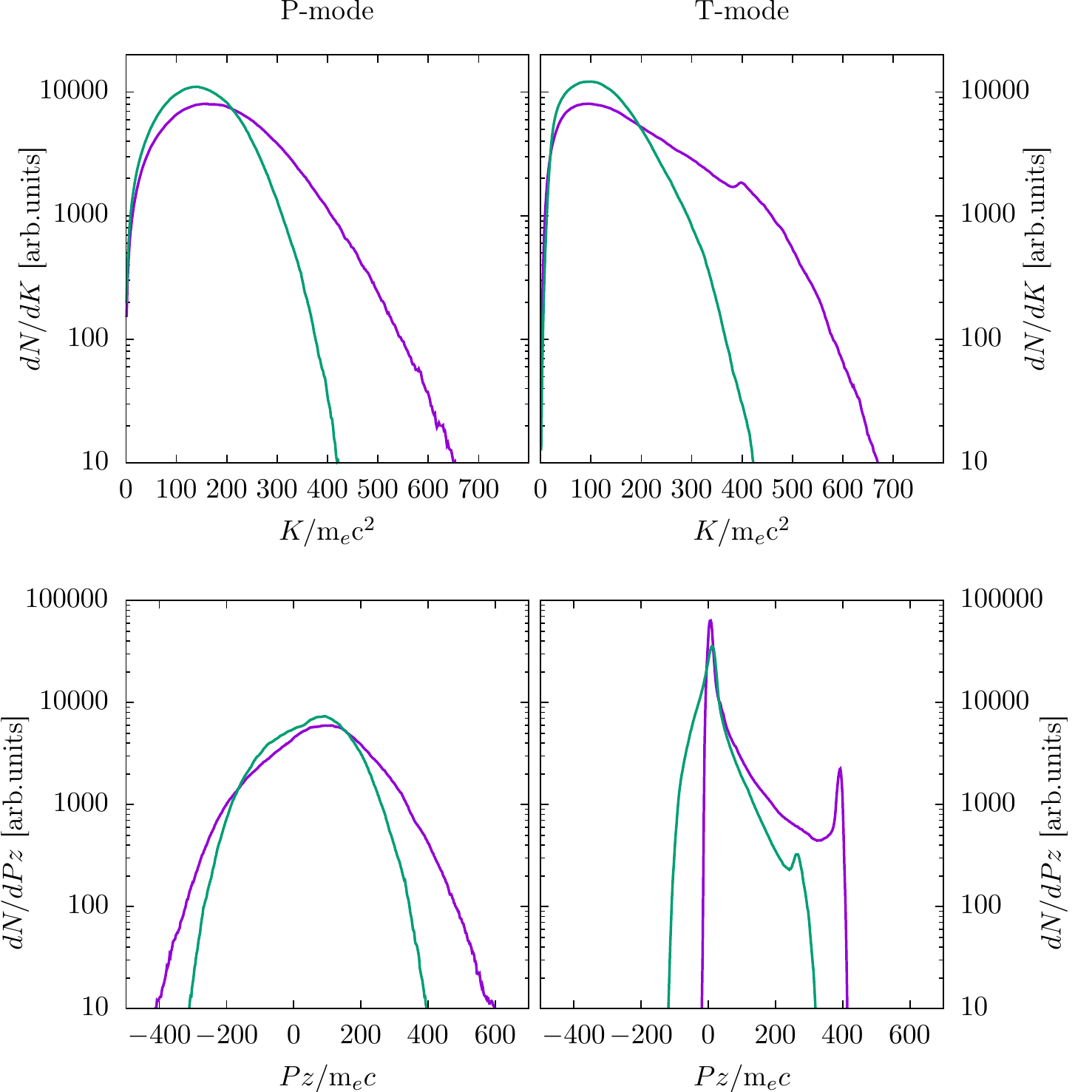}
 \put(10,98){$(a)$}
 \put(49,98){$(b)$}
 \put(10,48){$(c)$}
 \put(49,48){$(d)$}
 \end{overpic}
  \caption{2D simulation: Plots of the total kinetic energy spectrum at $t = 200 \, T_p$ for both $P$- ($(a)$ frame) and $T$-modes ($(b)$ frame). Plots of total $p_z$ spectrum at $t = 200 \, T_p$ for both $P$- ($(c)$ frame) and $T$-modes ($(d)$ frame). Green and purple lines refer to simulations with and without the inclusion of RF, respectively.}
  \label{fig:Enrr_2D}
\end{figure}

\begin{figure}
\includegraphics[width=0.5\textwidth]{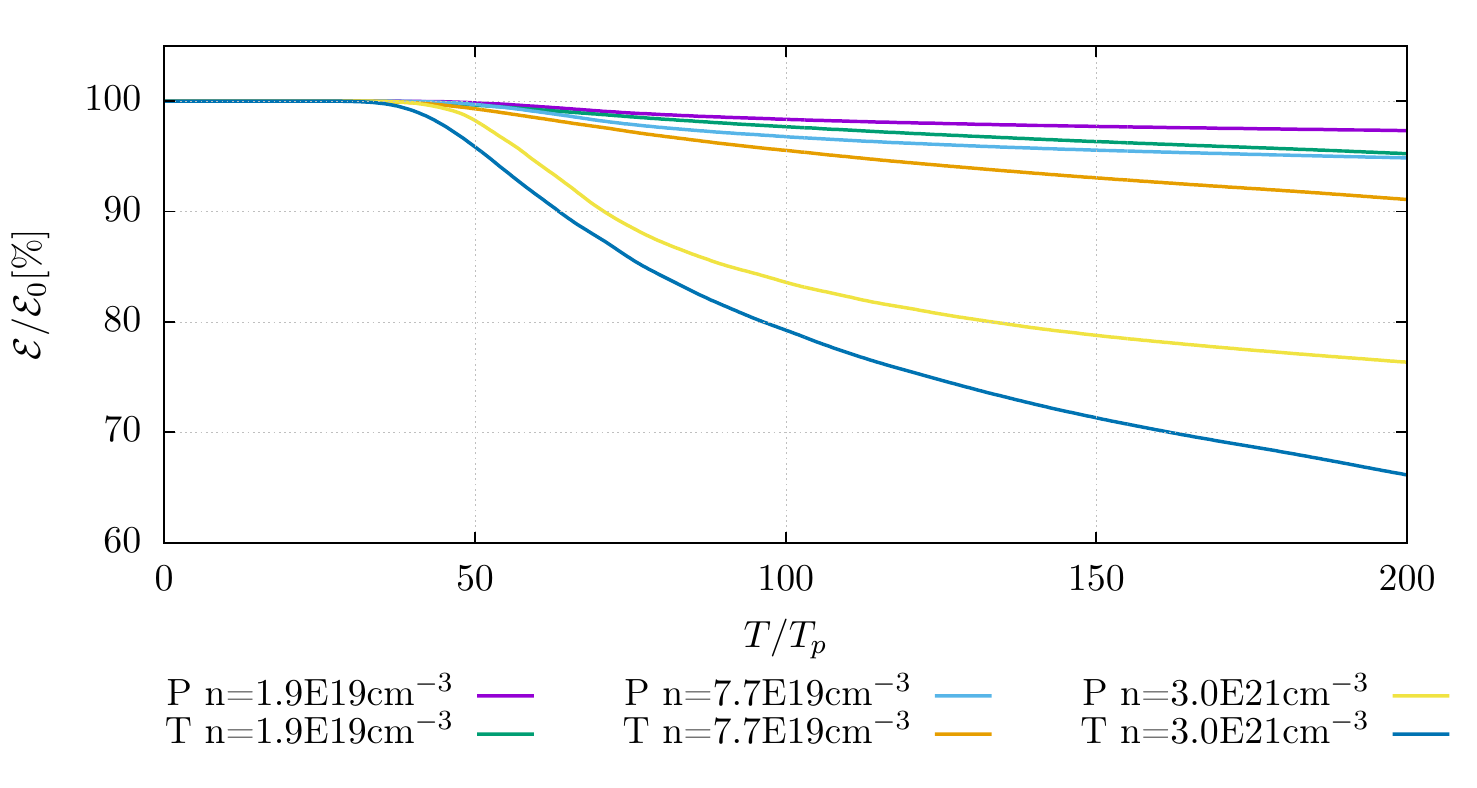}
  \caption{Dependence on time of the total energy in the system (plasma kinetic energy $+$ e.m. energy) normalized to the initial energy for a 1D simulations (red line) and for a 2D simulations (green line).}
\label{fig:etotrr}
\end{figure}

In this section we present the 2D simulations, comparing the results of $T$-mode and $P$-mode geometry. As already mentioned in Sec.\ref{setup}, two different codes have been used for numerical reasons. The rise of the instability is shifted in time between $T$- and $P$-mode simulations because of the different noise level in the two codes, although the growth rate is identical in benchmark cases. Thus, to make comparisons at the same physical time, the simulation time has been shifted in order that the instant at which the magnetic energy reaches its peak (marking the end of the linear growth stage) coincides for the two cases. \newline  
Fig.\ref{fig:J2DPT} shows the distribution of $J_z$ (the current density parallel to the beams direction) for the $P$- and $T-$plane cases, at three different times. In order to also represent the magnetic field distribution, Fig.\ref{fig:B2DPT} shows $B_z$ in the $P$-plane and $(B_x^2+B_y^2)^{1/2}$ in the $T$-plane at the same times of Fig.\ref{fig:J2DPT}.  \newline
In the early stage of the simulations ($t \lesssim 50$), the $P$-case shows parallel current filaments which are elongated in the beam direction $\hat{\bf z}$ and have almost the same width as the transverse structures in the $T$-case. This confirms that in the linear stage the most unstable wavevector is along the direction of the beams, i.e. the FI is of transverse nature. During this early stage, the amplitude of the field grows exponentially. At saturation ($t \simeq 55$), the beam energy converted into magnetic energy for the $T$-case is nearly two times the value for the $P$-case (Fig.\ref{fig:energy2D}), in agreement with an approximate energy equipartition. \newline
At later times ($t>55$) merging of small-scale filaments is observed in both the $T$- and $P$-cases, eventually leading at long times ($t>100$) to the formation of structures with a size close to that of the numerical box for both cases. However, significant differences are apparent between the $T$- and $P$-cases.\newline
For the $T$-case, the current distribution across a large-scale 2D structure is similar to that observed in 1D: the current peaks near the boundary of the island (which corresponds to the ``horned'' 1D profiles in Fig.\ref{fig:ByJz70000}) and has much weaker values well inside the island; locally, small scale filaments where the current changes sign are also observed, which are caused by the different orientation of the wave-vector with respect to the initial direction of the beams. The magnetic field is strongly localized along the boundary of the current structure, i.e. along null lines of $J_z$. The spatial correlation between the density of each species and the fields is also similar to that observed in 1D. The current and density distributions during the non-linear phase are similar to those observed as asymptotic numerical solutions of 2D Navier-Stokes and magnetohydrodynamic equations \citep{hossainJPP83}. The large scale structures of the magnetic field evolve slowly both in the shape and in the amplitude of the field for $t>100$. \newline
The distributions of total kinetic energy and $p_z$ for the $T$-case are shown in Fig.~\ref{fig:Enrr_2D} (frame $(b)$ and $(d)$). As in the 1D case, a peak at the upper cut-off $p_z \simeq 2 \, p_0 = 400 \, m_ec$ forms (see Fig.~\ref{fig:Enrr_2D}~(d)), while the spectral peak in the energy distribution (Fig.~\ref{fig:Enrr_2D}~(b)) disappears, because the energy in the tail of the distribution is ``smoothed'' out over the additional degree of freedom. Hence we can expect that the spectrum would be smeared out in a 3D case. While the inclusion of RF is found not to change the growth and development of the filaments significantly, it has a major impact on the high energy tail of the spectrum, reducing the cut-off by a factor of $\sim 2$, similarly to what observed in 1D. The amount of energy lost to radiation exceeds 30\% at the end of the simulation (Fig.\ref{fig:etotrr}).\newline
In the $P$-case, bending and tearing of filaments during the merging stage is observed. This leads to the generation of structures which are not homogeneous along the direction of the beams, i.e. to a spectrum of modes with $k_z \neq 0$. The latter processes can not be simply viewed as the growth of an unstable longitudinal mode: a Fourier analysis highlights a broad spectrum in $k_z$ at late times. Correspondingly, electrostatic fields are generated leading to breaking of the symmetry properties of the system for purely transverse EM perturbations. \newline
In the $P$-case the large scale structures of the magnetic field are less regular than in the $T$-case, showing a small-scale irregular structure at $t=200$. The decay of the magnetic field is much more pronounced with the magnetic energy becoming of the order of the electrostatic energy at the end of the simulation ($t=200$). This behavior is likely to be due to electrostatic fields causing heating of electrons and positrons in the simulations plane. The energy spectrum in the $P$-case becomes broader than in the $T$-case, with an higher energy cut-off. No narrow peak is observed at $2p_0$ in the $P$-case, confirming that peak formation is related to the conservation of canonical momentum along the direction of the beams in the $T$-case.  \newline
Finally, we discuss the importance of radiative losses due to the inclusion of RF for different geometry and density. Figure~\ref{fig:etotrr} reports the time evolution of the total energy (particle energy plus field energy) with respect to the initial kinetic energy of the beams, for different simulations with RF included. Radiative losses  are higher for the $T$-mode than for the $P$-mode, consistently with the higher fields generated in the $T$-mode case. The effect of RF is also stronger for higher densities $n_T$, which is also consistent with the magnetic field at saturation being proportional to $n_T^{1/2}$. At densities of the order of $10^{19} \mathrm{cm^{-3}}$, radiative losses reach a few per cent of the initial energy at the end of the simulation.  \newline
Although for very high density there is a major loss of energy due to the RF effect, the instability dynamics is not strongly changed with respect to the case without RF. The system organizes itself in filamentary structures for the current density which have almost the same size and features of the filaments obtained in the non-RF simulation. This behavior can be simply understood by noticing that the EM fields have to grow in order for RF to be important, so that the RF plays little role before the saturation phase. 
Moreover, in the ultra-relativistic case the dominant term of RF force (see \citet{landauRR}) is $\propto \gamma^2$. Thus the RF contribution is strongly increased by the acceleration of some particles to higher energy, which is maximized at the instability saturation stage. This is consistent with RF effects being more evident in the particle spectra, as shown in Fig.\ref{fig:Enrr_2D}.
%
%


\section{Conclusions}

In this work we have studied the evolution of the filamentation instability produced by two counter-streaming pair plasmas using PIC simulations in 1D and in 2D for both $T$- and $P$-modes, with and without radiation friction effects. The saturation level of the instability and the particle spectra are significantly different between $T$- and $P$-modes. In the $T$-mode case, the magnetic field at saturation is stronger and has a slower decay in time after reaching its maximum value; the particle spectrum shows the formation of a spectral peak at cut-off for which a simple theory has been presented. In the $P$-mode case, the magnetic field has a lower maximum value and has a faster decay, so that the magnetic energy becomes comparable to the electrostatic energy at the end of the simulations; the energy spectra show no peak but a higher energy cut-off. Radiation friction effects have been found to be strong only for relatively high density ($\sim 10^{20} \mathrm{cm^{-3}}$) and to modify strongly the particle spectra, cooling down the distribution functions and removing the highest energy particles, while the instability development is weakly affected. 
\newline

\section*{Acknowledgments}

We thank the National Center for Research and Development into the field of Information Technologies of the Italian Institute for Nuclear Physics (INFN-CNAF) for the technical assistance provided in running PICCANTE on the local cluster.
We thank M.~Vietri (Scuola Normale Superiore, Italy) for useful discussions. 

\bibliographystyle{mnras}
\bibliography{biblio}

\begin{thebibliography}{}
\makeatletter
\relax
\def\mn@urlcharsother{\let\do\@makeother \do\$\do\&\do\#\do\^\do\_\do\%\do\~}
\def\mn@doi{\begingroup\mn@urlcharsother \@ifnextchar [ {\mn@doi@}
  {\mn@doi@[]}}
\def\mn@doi@[#1]#2{\def\@tempa{#1}\ifx\@tempa\@empty \href
  {http://dx.doi.org/#2} {doi:#2}\else \href {http://dx.doi.org/#2} {#1}\fi
  \endgroup}
\def\mn@eprint#1#2{\mn@eprint@#1:#2::\@nil}
\def\mn@eprint@arXiv#1{\href {http://arxiv.org/abs/#1} {{\tt arXiv:#1}}}
\def\mn@eprint@dblp#1{\href {http://dblp.uni-trier.de/rec/bibtex/#1.xml}
  {dblp:#1}}
\def\mn@eprint@#1:#2:#3:#4\@nil{\def\@tempa {#1}\def\@tempb {#2}\def\@tempc
  {#3}\ifx \@tempc \@empty \let \@tempc \@tempb \let \@tempb \@tempa \fi \ifx
  \@tempb \@empty \def\@tempb {arXiv}\fi \@ifundefined
  {mn@eprint@\@tempb}{\@tempb:\@tempc}{\expandafter \expandafter \csname
  mn@eprint@\@tempb\endcsname \expandafter{\@tempc}}}

\bibitem[\protect\citeauthoryear{Amano \& Hoshino}{Amano \&
  Hoshino}{2009}]{amanoPoP09}
Amano T.,  Hoshino M.,  2009, \mn@doi [Physics of Plasmas]
  {http://dx.doi.org/10.1063/1.3240336}, 16,

\bibitem[\protect\citeauthoryear{Begelman, Blandford  \& Rees}{Begelman
  et~al.}{1984}]{begelmanRMP84}
Begelman M.~C.,  Blandford R.~D.,   Rees M.~J.,  1984, \mn@doi [Rev. Mod.
  Phys.] {10.1103/RevModPhys.56.255}, 56, 255

\bibitem[\protect\citeauthoryear{Bell}{Bell}{1978a}]{bellMNRAS78I}
Bell A.~R.,  1978a, \mn@doi [Monthly Notices of the Royal Astronomical Society]
  {10.1093/mnras/182.2.147}, 182, 147

\bibitem[\protect\citeauthoryear{Bell}{Bell}{1978b}]{bellMNRAS78II}
Bell A.~R.,  1978b, \mn@doi [Monthly Notices of the Royal Astronomical Society]
  {10.1093/mnras/182.3.443}, 182, 443

\bibitem[\protect\citeauthoryear{Blandford \& Eichler}{Blandford \&
  Eichler}{1987}]{blandfordPR87}
Blandford R.,  Eichler D.,  1987, \mn@doi [Physics Reports]
  {http://dx.doi.org/10.1016/0370-1573(87)90134-7}, 154, 1

\bibitem[\protect\citeauthoryear{Blandford \& Ostriker}{Blandford \&
  Ostriker}{1978}]{blandfordApJ78}
Blandford R.,  Ostriker J.,  1978, \mn@doi [Astrophysical Journal Letter]
  {http://dx.doi.org/10.1086/182658}, 221, L29

\bibitem[\protect\citeauthoryear{Blasi}{Blasi}{2013}]{blasiAAPR13}
Blasi P.,  2013, \mn@doi [The Astronomy and Astrophysics Review]
  {10.1007/s00159-013-0070-7}, 21

\bibitem[\protect\citeauthoryear{Blasi \& Amato}{Blasi \&
  Amato}{2011}]{blasi-amato11}
Blasi P.,  Amato E.,  2011, in Torres D.~F.,  Rea N.,  eds, Astrophysics and
  Space Science Proceedings, High-Energy Emission from Pulsars and their
  Systems.
Springer Berlin Heidelberg, pp 623--641, \mn@doi{10.1007/978-3-642-17251-9_50},
  \url {http://dx.doi.org/10.1007/978-3-642-17251-9_50}

\bibitem[\protect\citeauthoryear{Bret, Firpo  \& Deutsch}{Bret
  et~al.}{2004}]{bretPRE04}
Bret A.,  Firpo M.-C.,   Deutsch C.,  2004, \mn@doi [Phys. Rev. E]
  {10.1103/PhysRevE.70.046401}, 70, 046401

\bibitem[\protect\citeauthoryear{Bret, Gremillet  \& Dieckmann}{Bret
  et~al.}{2010}]{bretPOP10}
Bret A.,  Gremillet L.,   Dieckmann M.~E.,  2010, \mn@doi [Physics of Plasmas]
  {10.1063/1.3514586}, 17, 120501

\bibitem[\protect\citeauthoryear{Bret, Stockem, Fiuza, Ruyer, Gremillet,
  Narayan  \& Silva}{Bret et~al.}{2013}]{bretPoP13}
Bret A.,  Stockem A.,  Fiuza F.,  Ruyer C.,  Gremillet L.,  Narayan R.,   Silva
  L.~O.,  2013, \mn@doi [Physics of Plasmas (1994-present)]
  {http://dx.doi.org/10.1063/1.4798541}, 20,

\bibitem[\protect\citeauthoryear{Califano, Pegoraro  \& Bulanov}{Califano
  et~al.}{1997}]{califanoPRE97}
Califano F.,  Pegoraro F.,   Bulanov S.~V.,  1997, \mn@doi [Phys. Rev. E]
  {10.1103/PhysRevE.56.963}, 56, 963

\bibitem[\protect\citeauthoryear{Cerutti, Werner, Uzdensky  \&
  Begelman}{Cerutti et~al.}{2013}]{ceruttiApJ13}
Cerutti B.,  Werner G.~R.,  Uzdensky D.~A.,   Begelman M.~C.,  2013, The
  Astrophysical Journal, 770, 147

\bibitem[\protect\citeauthoryear{Chang, Spitkovsky  \& Arons}{Chang
  et~al.}{2008}]{changAJ08}
Chang P.,  Spitkovsky A.,   Arons J.,  2008, The Astrophysical Journal, 674,
  378

\bibitem[\protect\citeauthoryear{Di~Piazza, M\"uller, Hatsagortsyan  \&
  Keitel}{Di~Piazza et~al.}{2012}]{dipiazzaRMP12}
Di~Piazza A.,  M\"uller C.,  Hatsagortsyan K.~Z.,   Keitel C.~H.,  2012,
  \mn@doi [Rev. Mod. Phys.] {10.1103/RevModPhys.84.1177}, 84, 1177

\bibitem[\protect\citeauthoryear{Greenwood, Cartwright, Luginsland  \&
  Baca}{Greenwood et~al.}{2004}]{greenwoodJCP04}
Greenwood A.~D.,  Cartwright K.~L.,  Luginsland J.~W.,   Baca E.~A.,  2004,
  \mn@doi [Journal of Computational Physics]
  {http://dx.doi.org/10.1016/j.jcp.2004.06.021}, 201, 665

\bibitem[\protect\citeauthoryear{Hoshino \& Shimada}{Hoshino \&
  Shimada}{2002}]{hoshinoApJ02}
Hoshino M.,  Shimada N.,  2002, The Astrophysical Journal, 572, 880

\bibitem[\protect\citeauthoryear{Hossain, Matthaeus  \& Montgomery}{Hossain
  et~al.}{1983}]{hossainJPP83}
Hossain M.,  Matthaeus W.~H.,   Montgomery D.,  1983, \mn@doi [Journal of
  Plasma Physics] {10.1017/S0022377800001306}, 30, 479

\bibitem[\protect\citeauthoryear{Jaroschek \& Hoshino}{Jaroschek \&
  Hoshino}{2009}]{jaroschekPRL09}
Jaroschek C.~H.,  Hoshino M.,  2009, \mn@doi [Phys. Rev. Lett.]
  {10.1103/PhysRevLett.103.075002}, 103, 075002

\bibitem[\protect\citeauthoryear{Jaroschek, Lesch  \& Treumann}{Jaroschek
  et~al.}{2005}]{jaroschekApJ05}
Jaroschek C.~H.,  Lesch H.,   Treumann R.~A.,  2005, The Astrophysical Journal,
  618, 822

\bibitem[\protect\citeauthoryear{Kazimura, Sakai, Neubert  \& Bulanov}{Kazimura
  et~al.}{1998}]{kazimuraAJL98}
Kazimura Y.,  Sakai J.~I.,  Neubert T.,   Bulanov S.~V.,  1998, The
  Astrophysical Journal Letters, 498, L183

\bibitem[\protect\citeauthoryear{Landau \& Lifshitz}{Landau \&
  Lifshitz}{1975}]{landauRR}
Landau L.,  Lifshitz E.,  1975, The Classical Theory of Fields.
Butterworth-Heinemann

\bibitem[\protect\citeauthoryear{{Lemoine}, {Pelletier}, {Gremillet}  \&
  {Plotnikov}}{{Lemoine} et~al.}{2014}]{lemoineMnras14}
{Lemoine} M.,  {Pelletier} G.,  {Gremillet} L.,   {Plotnikov} I.,  2014,
  \mn@doi [mnras] {10.1093/mnras/stu213}, \href
  {http://adsabs.harvard.edu/abs/2014MNRAS.440.1365L} {440, 1365}

\bibitem[\protect\citeauthoryear{Liang, Boettcher  \& Smith}{Liang
  et~al.}{2013a}]{liangApJL13a}
Liang E.,  Boettcher M.,   Smith I.,  {2013}a, The Astrophysical Journal
  Letters, 766, L19

\bibitem[\protect\citeauthoryear{Liang, Fu, Boettcher, Smith  \&
  Roustazadeh}{Liang et~al.}{2013b}]{liangApJL13b}
Liang E.,  Fu W.,  Boettcher M.,  Smith I.,   Roustazadeh P.,  {2013}b, \mn@doi
  [The Astrophysical Journal Letters] {10.1088/2041-8205/779/2/L27}, 779, L27

\bibitem[\protect\citeauthoryear{Malkov \& Drury}{Malkov \&
  Drury}{2001}]{malkovRPP01}
Malkov M.~A.,  Drury L.~O.,  2001, Reports on Progress in Physics, 64, 429

\bibitem[\protect\citeauthoryear{Nishikawa et~al.,}{Nishikawa
  et~al.}{2009}]{nishikawaAPJ09}
Nishikawa K.-I.,  et~al., 2009, The Astrophysical Journal Letters, 698, L10

\bibitem[\protect\citeauthoryear{Piran}{Piran}{2005}]{piranRMP05}
Piran T.,  2005, \mn@doi [Rev. Mod. Phys.] {10.1103/RevModPhys.76.1143}, 76,
  1143

\bibitem[\protect\citeauthoryear{Sgattoni, Fedeli  \& Sinigardi}{Sgattoni
  et~al.}{2014}]{PICCANTE}
Sgattoni A.,  Fedeli L.,   Sinigardi S.,  2014, ``PICCANTE, an open-source
  massively parallel Particle-In-Cell code'',
  \texttt{http://aladyn.github.io/piccante/}

\bibitem[\protect\citeauthoryear{{Sgattoni}, {Fedeli}, {Sinigardi},
  {Marocchino}, {Macchi}, {Weinberg}  \& {Karmakar}}{{Sgattoni}
  et~al.}{2015}]{PICCANTE2}
{Sgattoni} A.,  {Fedeli} L.,  {Sinigardi} S.,  {Marocchino} A.,  {Macchi} A.,
  {Weinberg} V.,   {Karmakar} A.,  2015, preprint, \href
  {http://adsabs.harvard.edu/abs/2015arXiv150302464S} {} (\mn@eprint {arXiv}
  {1503.02464})

\bibitem[\protect\citeauthoryear{Silva, Fonseca, Tonge, Dawson, Mori  \&
  Medvedev}{Silva et~al.}{2003}]{silvaApJ03}
Silva L.~O.,  Fonseca R.~A.,  Tonge J.~W.,  Dawson J.~M.,  Mori W.~B.,
  Medvedev M.~V.,  2003, The Astrophysical Journal Letters, 596, L121

\bibitem[\protect\citeauthoryear{Spitkovsky}{Spitkovsky}{2008}]{spitkovskyAJL08}
Spitkovsky A.,  2008, The Astrophysical Journal Letters, 682, L5

\bibitem[\protect\citeauthoryear{Tamburini, Pegoraro, Piazza, Keitel  \&
  Macchi}{Tamburini et~al.}{2010}]{tamburiniNJP10}
Tamburini M.,  Pegoraro F.,  Piazza A.~D.,  Keitel C.~H.,   Macchi A.,  2010,
  New Journal of Physics, 12, 123005

\bibitem[\protect\citeauthoryear{{Vranic}, {Martins}, {Fonseca}  \&
  {Silva}}{{Vranic} et~al.}{2015}]{vranicXXX15}
{Vranic} M.,  {Martins} J.~L.,  {Fonseca} R.~A.,   {Silva} L.~O.,  2015,
  preprint, \href {http://adsabs.harvard.edu/abs/2015arXiv150202432V} {}
  (\mn@eprint {arXiv} {1502.02432})

\makeatother
\end{thebibliography}

\end{document}